\documentclass[aps,prl,twocolumn,showpacs,floatfix]{revtex4}
\usepackage{graphics}
\usepackage{dcolumn}
\usepackage{bm}

\begin{document} 
\title{Particle Physics Implications of a Recent Test of the 
Gravitational Inverse-Square Law}

\author{E.G. Adelberger}
\author{B.R. Heckel}
\author{S. Hoedl}
\author{C.D. Hoyle}\altaffiliation[Present address: ]{Department of Physics and Astronomy, Humboldt State University, Arcata CA, 95521-8299}
\author{D.J. Kapner}\altaffiliation[Present address: ]{Kavli Institute for Cosmological Physics, University of Chicago, Chicago IL, 60637}
\affiliation{Center for Experimental Nuclear Physics and Astrophysics, Box 354290,
University of
Washington, Seattle, WA 98195-4290}
\author{A. Upadhye}
\affiliation{Department of Physics, Princeton University, Princeton, NJ 08544}

\date{\today}
\begin{abstract}
We use data from our recent search for violations of the gravitational inverse-square law to
constrain dilaton, radion and chameleon exchange forces as well as arbitrary vector or scalar Yukawa interactions. We test the interpretation of the PVLAS effect and a conjectured ``fat graviton'' scenario and constrain the $\gamma_5$ couplings of pseuodscalar bosons and arbitrary power-law interactions. 
\end{abstract}
\maketitle
In a recent Letter\cite{ka:06}, we reported a sensitive torsion-balance search search for Yukawa violations of the gravitational inverse-square law (ISL) of the form
\begin{equation}
V_{ab}(r)=-\alpha G \frac{M_a M_b}{r}\exp(-r/\lambda)~.
\end{equation}
However, space limitations prevented us from discussing some implications of that result and constraining other forms of possible breakdowns of the ISL. 
In this Letter we use the data from Ref.~\cite{ka:06} to obtain upper bounds on several interesting exotic interactions.%
\section {Yukawa interactions from generic scalar or vector boson exchange}
Exchange of scalar or vector bosons, $\phi$, with mass $m$ between two non-relativisitic fermions generically produces a potential
\begin{equation}
V_{ab}(r)= \mp\frac{g_{S,V}^a g_{S,V}^b}{4\pi r} {\exp (-r/\lambda)}~,
\label{eq: vector}
\end{equation}
where the $-$ and $+$ signs refer to scalar and vector interactions respectively, 
and $\lambda=\hbar/mc$. For arbitrary vector interactions between electrically neutral atoms with proton and neutron numbers $Z$ and $N$, we have
\begin{equation}
g_V =g_V^0 (Z \cos \tilde{\psi}  + N\sin \tilde{\psi})~,
\label{eq: gV}
\end{equation}
where 
\begin{equation}
\tilde{\psi}\equiv \arctan\frac{\tilde{q}^n_V}{\tilde{q}^p_V+\tilde{q}^e_V}
\label{eq: psi}
\end{equation} 
$\psi$ is an angle that, in principle, could have any value between $-\pi/2$ and $\pi/2$, and the $\tilde{q}_V$'s are vector ``charges''.  Eqs.~\ref{eq: gV} and \ref{eq: psi} can also be applied to scalar interactions, even though they are not exact because scalar charges are not conserved and binding energy can carry a charge.

Expressing Eq.~\ref{eq: vector} in terms of Eq.~1,
we have
\begin{equation}
\frac{g^0_a g^0_b}{4\pi} = \alpha G u^2 \left( \left[\frac{\tilde{q}}{\mu}\right]_a \left[\frac{\tilde{q}}{\mu}\right]_b \right)^{-1}~,
\label{eq: gsquared}
\end{equation}
where $\mu=M/u$ with $M$ and $u$ being the atomic mass and atomic mass unit respectively, and
\begin{equation}
\left[\frac{\tilde{q}}{\mu}\right]= \left[\frac{Z}{\mu}\right] \cos \tilde{\psi} + \left[\frac{N}{\mu}\right] \sin{\tilde{\psi}}~.
\label{eq: charge}
\end{equation}
The molybdenum pendulum and attractor used in Ref.~\cite{ka:06} have
$\left[ Z/\mu \right]\!=0.4378134$ and $\left[ N/\mu \right]\!=0.5631686$.
%
%
\begin{figure}
\hfil\scalebox{.49}{\includegraphics*[60pt,40pt][574pt,440pt]{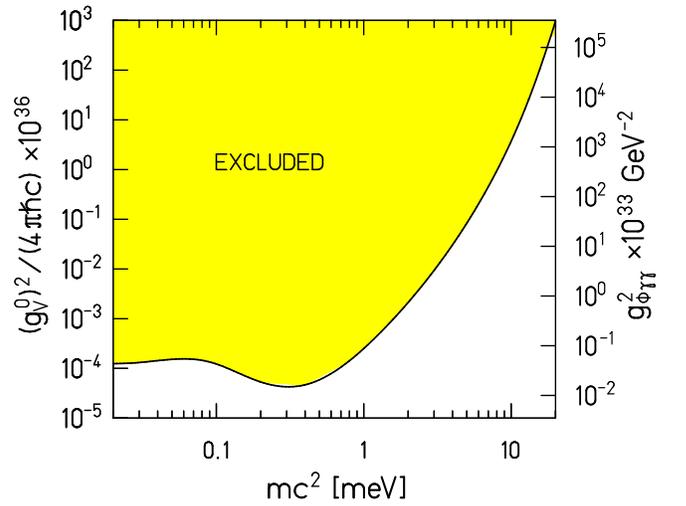}}\hfil
\caption{(color online) 95\% confidence constraints on scalar or vector Yukawa interactions. Left vertical scale: vector interactions coupled to $B-L$ ({\em i.e.} $\tilde{\psi} = \pi/2$); right vertical scale: scalar $\phi \gamma\gamma$ couplings inferred as discussed below.}
\label{fig: alphalambda}
\end{figure}
Figure~\ref{fig: alphalambda} illustrates the upper limits implied by the results of Ref.~\cite{ka:06} on vector interactions coupled to $B-L$ where $B$ and $L$ are baryon and lepton
numbers, respectively.
%
%
\begin{figure}
\hfil\scalebox{.47}{\includegraphics*[60pt,40pt][503pt,430pt]{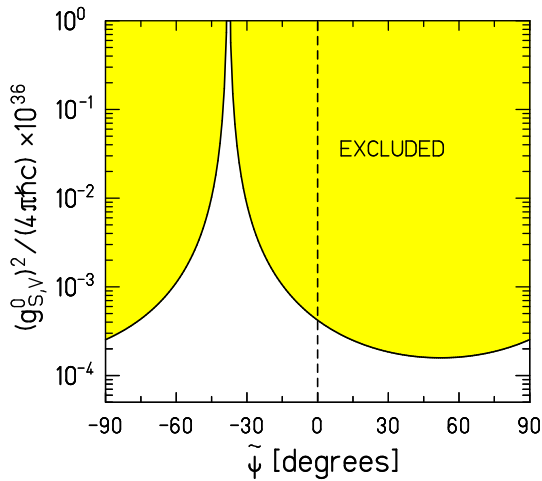}}\hfil
\caption{(color online) 95\% confidence constraints on couplings of a boson of mass $mc^2=1$ meV as a function of the charge parameter $\tilde{\psi}$.
Constraints for other values of $m$ can be found by using Fig.~\ref{fig: alphalambda} to scale the couplings as a function of $m$.}
\label{fig: theta5}
\end{figure}
Figure~\ref{fig: theta5} shows how the upper limits on $g_{S,V}^2$ depend on the parameter $\tilde{\psi}$
that specifies the charge $\tilde{q}$.

\section {Yukawa interactions from radion and dilaton exchange.}
In string theories, the geometry of spacetime is expected to be dynamical with the radii of new dimensions fluctuating independently at each point in our 4-dimensional spacetime. In an effective low-energy theory, the volume of the extra dimensions must be stabilized by {\em radions}, low-mass spin-0 fields with gravitational-strength couplings that determine the radius of the $n$ ``large'' extra dimensions. Radion exchange will produce a Yukawa force with a strength and range\cite{ad:03}
\begin{equation}
\alpha=\frac{n}{n+2}
\label{eq:radion}
\end{equation}
\begin{equation}
\lambda \sim \sqrt{\frac{\hbar^3}{c G M_{\ast}^4}} \approx 2.4 \left[ \frac{1~{\rm TeV}}{M_{\ast} c^2} \right]^2\;{\rm mm}~,
\label{eq:radion range}
\end{equation}
where $M_{\ast}$ is the unification mass.
In many cases the radion-mediated force is the longest-range effect of new dimensions\cite{an:98} because it
does not diminish as the number of new dimensions increases.
For $n=1$ and $n=6$ ($\alpha=1/3$ and $\alpha=3/4$), the data of Ref.~\cite{ka:06} give
\begin{eqnarray}
M_{\ast}(n=1)&\geq& 5.7~{\rm Tev}/c^2 \\
M_{\ast}(n=6)&\geq& 6.4 ~{\rm Tev}/c^2~.
\end{eqnarray}
String theories predict a scalar partner of the graviton, the dilaton, whose mass is initially zero. 
Equivalence Principle experiments have ruled out
a massless dilaton. Kaplan and Wise\cite{ka:00} evaluated the coupling of a low-mass dilaton to
strongly interacting matter and shown that its coupling to matter should satisfy
$1 \leq \alpha \leq 1000$. Figure~6 of Ref.~\cite{ka:06} then sets a 95\% confidence lower bound
on the dilaton mass of  
\begin{equation}
m c^2 \geq  3.5~{\rm meV}~.
\end{equation}

\section{Is the PVLAS effect evidence for new physics?}
\begin{figure}[b]
\hfil\scalebox{.35}{\includegraphics*[10pt,1pt][626pt,150pt]{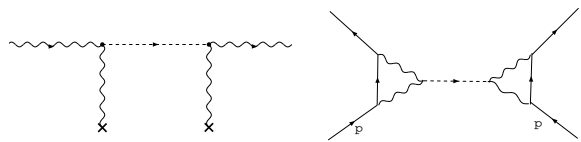}}\hfil
\caption{Left: diagram of the process proposed to explain the PVLAS result\cite{za:06}.
Right: diagram of a related process that would give a signal in gravitational experiments. Wiggly and dashed lines are photons and scalar bosons, respectively.}
\label{fig: PVLAS diag}
\end{figure}

Recently, the PVLAS collaboration\cite{za:06} studied the propagation of optical photons through a vacuum containing a strong transverse $B$ field. They reported an optical rotation at least $10^4$ times larger than the QED prediction, and speculated that this was evidence for a new spin-zero particle that, through a second-order process, mixes with the photon in a magnetic field as shown in Fig.~\ref{fig: PVLAS diag}. The apparent sign of the observed rotation requires the new particle to be a scalar (as opposed to pseudoscalar) boson\cite{ca:06}, and the magnitude requires
\begin{eqnarray}
1.0 ~\rm{meV} &\leq& m_{\phi}c^2 \leq 1.5 ~\rm{meV} \nonumber \\
1.7\times 10^{-6} ~\rm{GeV}^{-1} &\leq& g_{\phi \gamma\gamma} \leq 5\times 10^{-6} ~\rm{GeV}^{-1}~.
\label{eq: PVLAS claim}
\end{eqnarray}

This $\phi \gamma\gamma$ vertex generates, by the second-order process shown in Fig.~\ref{fig: PVLAS diag}, an effective scalar interaction between two protons, which to leading order is estimated to be\cite{ma:06}  
\begin{equation}
\frac{g_S^p}{\sqrt{4 \pi \hbar c}} \sim g_{\phi \gamma\gamma} \left(\frac{\alpha}{\pi}m_p \right)~.
\label{eq: our 2 gamma bound}
\end{equation}
Assuming that $\tilde q_S^p$ is large compared to $\tilde{q}_S^e$ and $\tilde{q}_S^n$ ({\em i.e.} $\tilde{\psi}\approx 0$), our scalar constraints (Fig.~\ref{fig: theta5} with $\tilde{\psi}=0$) place the upper limit shown in
Fig.~\ref{fig: alphalambda}; in particular
\begin{equation}
g_{\phi \gamma\gamma} \alt 1.6 \times 10^{-17}~\rm{GeV}^{-1}~, 
\end{equation}
which is inconsistent with Eq.~\ref{eq: PVLAS claim} by a factor of
$\sim 10^{11}$. 

For $m_{\phi}c^2 \leq 20 ~\rm{meV}$, the bound in Eq.~\ref{eq: our 2 gamma bound} and Fig.~\ref{fig: alphalambda} improves on the astrophysical constraint\cite{ra:96} by a factor up to $10^8$. (Both of these bounds would be relaxed in models where the $\phi \gamma\gamma$ interaction has an additional low-energy form factor.)
%
%
\section {Yukawa interactions from chameleon exchange}
Chameleons are scalar fields that couple
to themselves and to matter with gravitational strength\cite{kh:04}. 
Chameleon exchange leads to an effective
potential density\cite{up:06} 
\begin{equation}
V_{\rm eff}(\phi,\vec{x}) = \frac{1}{2}m_\phi^2\phi^2+\frac{\gamma}{4!}\phi^4 -\frac{\beta}{M_{\rm Pl}}\rho(\vec{x})\phi~,
\end{equation}
where $\gamma$ characterizes the strength of the self interaction, $\beta$ characterizes the coupling of the scalar field to 
matter, and $M_{\rm Pl}$ is the reduced Planck mass. The ``natural'' values of $\beta$ and $\gamma$ are $\approx 1$.
In the presence of matter with density $\rho$, a massless chameleon field acquires an 
effective mass\cite{up:06},  
\begin{equation}
m_{\rm eff}(\rho)=\frac{\hbar}{c} \left(\frac{9}{2}\right)^{\!\!1/6}\gamma^{1/6}\left(\frac{\beta \rho}{M_{\rm Pl}}\right)^{\!\!1/3}~,
\end{equation} 
that dramatically weakens experimental constraints
because
only a small amount of material near the surface with thickness ${\cal O}(\hbar/(m_{\rm eff} c))$ 
contributes to a long-range force\cite{kh:04,up:06,fe:06,mo:06}. For $\rho= 10$~g/cm$^2$ and $\beta=\gamma=1$, this skin thickness is about 60~$\mu$m.

\begin{figure}[!]
\hfil\scalebox{.42}{\includegraphics*[70pt,35pt][588pt,526pt]{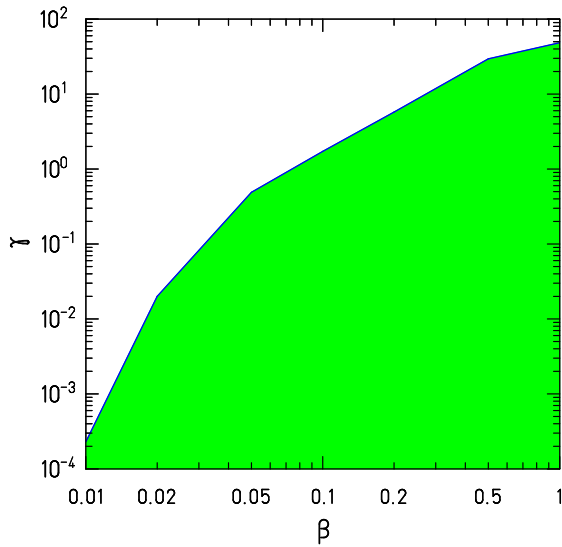}}\hfil
\caption{(color online) $2\sigma$ constraints on the chameleon parameter $\beta$ as a function of $\gamma$ from the data of Ref.~\cite{ka:06}.
The shaded area is ruled out at 95\% confidence. 
The chameleon signal is strongest when the chamelon length scale is comparable to the 1 mm hole thickness. For much larger length scales, the field varies little over the pendulum and attractor, giving a weak signal. 
For length
scales much smaller than the thickness of the BeCu foil, the signal is``screened" by the foil.}
\label{fig: betagamma}
\end{figure}

Using the method outlined in Ref.~\cite{up:06}, we calculated the 21$\omega$ chameleon torque, $N_{21}(\beta, \gamma,s)$, 
as a function of pendulum/attractor separation $s$ for the apparatus of Ref.~\cite{ka:06}. The combined Ref.~\cite{ka:06} data were fitted with the
Newtonian torque plus $N_{21}(\beta, \gamma,s)$ to generate the 95\% confidence level constraints on $\gamma$ as a function of $\beta$
shown in Fig.~\ref{fig: betagamma}. A substantial region of parameter space around the 
natural values $\beta \approx 1$ and $\gamma \approx 1$ is strongly excluded.
%
%
\section {The ``fat graviton'' conjecture}
Sundrum\cite{su:04} suggested a solution to the cosmological constant problem, namely that the observed ``dark energy'' density is much smaller than the vacuum energy density predicted by the usual rules of quantum mechanics. He conjectured that this could be explained if the graviton is a ``fat'' object whose size $\ell_g$ prevents it from ``seeing'' the short-distance physics that dominates the vacuum energy. In his conjecture, the gravitational force vanishes
at sufficiently small separations compared to $\ell_g$. 
We test this scenario by assuming that the gravitational force is
\begin{equation}
F_{12}^{\rm fat}(r)=-G\frac{M_1 M_2}{r^2}\left[1-\exp(-.914r/\ell_g)^3 \right]~,
\label{eq: fat graviton}
\end{equation}
which has a shape similar to the force shown pictorially in Fig.~8 of Ref.~\cite{su:04}; it vanishes at $r=0$ and has a maximum at $r=\ell_g$. Sundrum argues that naturalness requires $\ell_g \geq20~\mu$m. Our results 
require $\ell_g \leq 98~\mu$m at 95\% confidence.
Our upper limit on $\ell_g$ is larger than our $44~\mu$m limit on the  size of an extra dimension because our data probe the large-distance tail of the new potentials, and the ``fat-graviton'' force falls off much more rapidly with increasing separation than does a Yukawa force. 
\section {Power-law potentials and multi-particle exchange forces}
\label{sec:power law constraints} 
We constrained power-law interactions of the form
\begin{equation}
V_{ab}^k(r)=-G \frac{M_a M_b}{r}\beta_k \left( \frac{\rm 1~mm}{r} \right)^{k-1}
\label{eq: power law definition}
\end{equation}
by fitting the combined data of
Ref.~\cite{ka:06} with a function that contained the
Newtonian term and a single power-law term with $k=2$, 3, 4, or 5. The results are listed in Table~\ref{tab: power-law constraints} together with constraints from previous ISL tests.
%
\begin{table}[b]
\caption{68\% confidence laboratory constraints on power-law potentials 
from this work and previous\cite{ho:85,ho:04} results.}
\begin{ruledtabular}
\begin{tabular}{lcc}
$k$ &   $|\beta_k|$(this work) &  $|\beta_k|$(previous work)  \\
\hline
2   &   $4.5 \times 10^{-4}$   &  $1.3 \times 10^{-3}$\cite{ho:85}  \\  
3   &   $1.3 \times 10^{-4}$   &  $2.8 \times 10^{-3}$\cite{ho:04}  \\ 
4   &   $4.9 \times 10^{-5}$   &  $2.9 \times 10^{-3}$\cite{ho:04}  \\ 
5   &   $1.5 \times 10^{-5}$   &  $2.3 \times 10^{-3}$\cite{ho:04}  \\ 
\end{tabular}
\end{ruledtabular}
\label{tab: power-law constraints}
\end{table}

Power-law interactions arise from higher-order exchange processes with simultaneous exchange of multiple massless bosons. Second-order
processes are particularly interesting when the exchanged particles are unnatural-parity bosons (for which the 1st-order force vanishes when averaged over unpolarized test bodies) or fermions (for which the 1st-order process is forbidden). 
Potentials with $k=2$ are generated by the simultaneous exchange of two massless scalar\cite{su:93}  bosons. Simultaneous exchange of 
massless pseudoscalar\cite{fe:98} particles between fermions $a$ and $b$ with $\gamma_5$-couplings to
$g_a$ and $g_b$, gives a $k=3$ potential 
\begin{equation}
V_{ab}(r) = -\frac{\hbar}{c^3}\frac{1}{64 \pi^3}\frac{(g_P^a g_P^b)^2}{M_a M_b} \frac{1}{r^3}~.
\label{eq: massless gamma-5}
\end{equation}

Potentials with $k=5$ are produced by the simultaneous exchange of two massless pseudoscalars with $\gamma_5\gamma_{\mu} \partial^{\mu}$ couplings such as axions or Goldstone bosons\cite{fe:98},
but in this case our constraints 
are not competitive with astrophysical bounds on the first-order process\cite{ra:96}.
\subsection{Constraints on $\gamma_5$-coupled pseudoscalars}
Our limits on $\beta_3$ in Table~\ref{tab: power-law constraints}, together with Eq.~\ref{eq: massless gamma-5}, constrain the $\gamma_5$ couplings of massless pseudoscalars to neutrons and protons. 
\begin{eqnarray}
\Gamma \!&\equiv&\! \left[ \frac{Z}{\mu}\right]^2 \!\! \frac{(g_P^p)^4}{(\hbar c)^2} \!+ \!\left[ \frac{N}{\mu}\right]^2 \! \frac{(g_P^n)^4}{(\hbar c)^2} 
\!+\! 2\left[ \frac{Z}{\mu}\right] \left[ \frac{N}{\mu}\right] \! \frac{(g_P^p)^2}{\hbar c}\frac{(g_P^n)^2}{\hbar c}  \nonumber \\ 
  &=& \beta_3 \frac{c G}{\hbar^3} 64 \pi^3 u^4 (1~{\rm mm})^2 = 2.56 \times 10^{-10} \beta_3~. 
\end{eqnarray}
Couplings to electrons can be ignored because of the very small upper limit on such couplings deduced from an electron-spin-dependence experiment (see Ref.~\cite{fi:99}). 
%
%
\begin{figure}[!]
\hfil\scalebox{.47}{\includegraphics*[60pt,36pt][520pt,420pt]{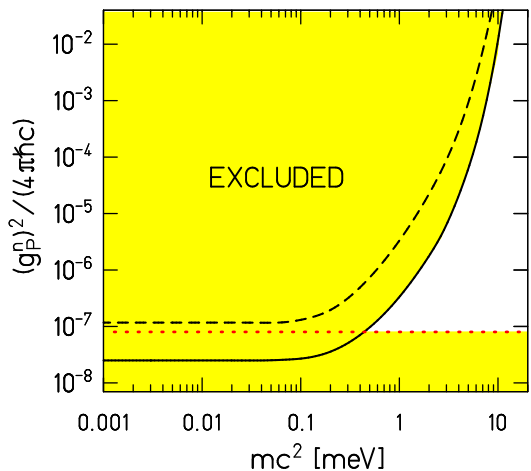}}\hfil
\caption{(color online) 68\%-confidence constraints on the $\gamma_5$ couplings of massive pseudoscalars to neutrons plotted against pseudoscalar mass $m$. The solid and dashed curves are from
the ISL tests of Refs.~\cite{ka:06} and \cite{ho:04} respectively. The horizontal dotted line shows the SN1987a constraint\cite{ra:96}.}
\label{fig: PS constraints}
\end{figure}
We constrained the $\gamma_5$ couplings of pseudoscalars of mass $m$ by fitting the Ref.~\cite{ka:06} data in terms of the Newtonian potential plus the appropriate generalization\cite{fe:99,ad:03a} of Eq.~\ref{eq: massless gamma-5},
\begin{equation}
V_{ab}(r) = -\frac{\hbar}{c^3}\frac{(g_P^a g_P^b)^2}{32 \pi^3 M_a M_b} \frac{K_1(2r/\lambda)}{\lambda r^2}~,
\label{eq: massive}
\end{equation}
where $K_1$ is a modified Bessel function and $\lambda=\hbar/(m c)$.
Our bounds on $g_P^n$ are shown in Fig.~\ref{fig: PS constraints}. The observed SN1987a  neutrino pulse excludes 
$8 \times 10^{-14} < (g_P^n)^2/(4\pi\hbar c) < 8 \times 10^{-8}$\cite{ra:96}. 
Stronger couplings are allowed because the 
pseudoscalars would have been trapped in the star;
we exclude this possibility for $\gamma_5$-coupled pseudoscalars with $mc^2 \leq 0.6~$meV.
Helioseismology constraints on exotic energy loss processes provide a 
limit $(g_P^p) ^2/(4\pi\hbar c) < 3\times10^{-9}$\cite{ra:06a}. 

We thank D.B. Kaplan, E. Mass\'o, A. Nelson and G. Raffelt for informative conversations. 
This work was supported by NSF Grant PHY0355012 and by the DOE Office of Science.

\end{document}